\documentclass[copyright,creativecommons]{eptcs}
\providecommand{\event}{ICE 2019} 
\usepackage{textcomp}
\usepackage{textgreek}
\usepackage{doi}
\usepackage[all]{foreign}

\usepackage{enumitem}
\setlist[enumerate]{nosep}

\usepackage{ulem}
\makeatletter
\def\uwave{\bgroup \markoverwith{\lower3.5\p@\hbox{\sixly \textcolor{red}{\char58}}}\ULon}
\font\sixly=lasy6 
\makeatother

\usepackage{inconsolata}
\usepackage{listings}
\usepackage{listings-rust}
\usepackage{cleveref}

\title{Rusty Variation\\ Deadlock-free Sessions with Failure in Rust}
\author{Wen Kokke
  \institute{LFCS, University of Edinburgh, 10 Crichton St, EH8 9AB, Edinburgh, United Kingdom}
  \email{wen.kokke@ed.ac.uk}
}
\def\titlerunning{Rusty Variation}
\def\authorrunning{Wen Kokke}

\begin{document}

\maketitle

\begin{abstract}
  Rusty Variation (RV) is a library for session-typed communication in Rust which offers strong compile-time correctness guarantees. Programs written using RV are guaranteed to respect a specified protocol, and are guaranteed to be free from deadlocks and races.
\end{abstract}

\section{Introduction}\label{sec:introduction}
In concurrent programming, processes commonly exchange information by sending messages on shared channels. However, this communication often does not follow a specified protocol, and if it does, the correctness of the implementation with respect to that protocol is rarely checked.

Session types are a substructural typing discipline, introduced by Honda~\cite{honda1993}, which describe communication protocols as types, capturing both the types and the order of messages. Session type systems, then, ensure that programs correctly implement specified protocols.

Rust is an immensely popular systems programming language.\footnote{\url{https://insights.stackoverflow.com/survey/2019/}} It aims to combine efficiency with abstraction and thread and memory safety. However, out of the box, it does not offer a way to specify protocols and verify communication.

In this paper, we present Rusty Variation\footnote{\url{https://github.com/wenkokke/rusty-variation/}} (RV), a library for session-typed communication in Rust which allows users to specify and check the adherence to communication protocols, in addition to offering strong correctness guarantees, such as freedom from deadlocks and unwanted races, while accounting for the possibility of failure. It is these strong guarantees of correctness, together with our account of failure, which sets us apart from previous work~\cite{jespersen2015}. We assume some familiarity with Rust. For an introduction to Rust, we refer the reader to \textit{The Rust Programming Language}~\cite{rust2019}.

We base our work on Exceptional GV by Fowler \etal~\cite[EGV]{fowler2019}. Exceptional GV is a descendant of Good Variation~\cite[GV]{wadler2012}, which adds support for exceptions and the cancellation of sessions. GV is a \textlambda-calculus with operations for forking threads and sending and receiving values. Crucially, EGV has an affine type system, and permits the explicit cancellation of sessions. EGV guarantees session fidelity, global progress, freedom from deadlocks and livelocks, confluence, and termination. That is to say, there are no locks or race conditions, and all programs eventually halt. For a discussion of EGV and its metatheory, we refer the reader to the work by Fowler \etal~\cite{fowler2019}.

Rusty Variation is an implementation of Exceptional GV (EGV) in Rust. We claim that RV is a faithful implementation of EGV, preserving much of EGV's metatheory, meaning RV guarantees session fidelity, global progress, confluence, and freedom from deadlocks and unwanted races. However, we depart from EGV in two ways:
First, EGV relies upon exceptions and handlers. Rust, however, rejects exceptions in favour of monadic errors, \ie, wrapping possibly failing results in error types such as \lstinline{Option<T>} and \lstinline{Result<T,E>}.
Second, the core language of EGV does not include general recursion. Rust does. In the presence of non-termination, we cannot guarantee livelock freedom, or---well---termination.

RV is currently implemented on top of the channels provided by the \lstinline{crossbeam_channel} crate, but it can be implemented using any library which offers one-shot channels and selection.

The paper proceeds as follows. In~\cref{sec:contribution}, we describe the programming interface offered by Rusty Variation, and provide several example programs. In~\cref{sec:related-work}, we compare Rusty Variation to the related work by Laumann \etal~\cite{jespersen2015}. Finally, in~\cref{sec:future-work}, we describe future work on verifying the correctness guarantees offered by RV.

\section{Rusty Variation}\label{sec:contribution}

\subsection{Types, sessions, and duality}
There are three basic types which we use to describe session protocols---or \emph{session types}: \lstinline{Send}, \lstinline{Recv}, and \lstinline{End}. We refer to values of these types as \emph{session endpoints} or, when referring to both endpoints, \emph{sessions}. Session endpoints wrap primitive channels---see~\cref{sec:implementation}. Any type built using these primitive types is a valid session type:
\begin{lstlisting}
pub struct Send<T, S: Session> { ... }
pub struct Recv<T, S: Session> { ... }
pub struct End { ... }
\end{lstlisting}
Session types are public, but their member functions are not. Therefore, users cannot construct their own misbehaving sessions. The continuation of a session must itself be a session. This is enforced by the \lstinline{Session} constraint. In addition to this ``kinding'' of session types, the \lstinline{Session} trait also requires any session type to implement duality and session generation:
\begin{lstlisting}
pub trait Session {
  type Dual: Session<Dual=Self>;
  fn new() -> (Self, Self::Dual);
}
\end{lstlisting}
Duality works as expected: \lstinline{Send<T, S>} is dual to \lstinline{Recv<T, S::Dual>}---and vice versa---and \lstinline{End} is self-dual. The constraint \lstinline{Session<Dual=Self>} enforces that the dual of a session type is a session type, and that \lstinline{S::Dual::Dual} is equal to \lstinline{S} (involutivity).

The \lstinline{new} function generates a new session-typed channel and returns two \emph{dual} channel endpoints. Unlike the rest of the functions discussed in this section, \lstinline{new} is \emph{not} part of the public interface---see \cref{sec:limitations}.

Using these types, we can define the types of servers which offer squaring or negation as follows, where \lstinline{i32} is the type of 32-bit integers:
\begin{lstlisting}
type SqrSrv = Recv<i32, Send<i32, End>>;
type NegSrv = Recv<i32, Send<i32, End>>;
\end{lstlisting}
The client types are obtained by duality.

\subsection{Send, receive, close, and fork}\label{sec:basics}
The \lstinline{send} and \lstinline{recv} functions send and receive values in sessions. Sending is non-blocking, and never fails. Sending a value on a cancelled channel is equivalent to cancelling that value---see \cref{sec:cancellation}. Receiving is blocking, and fails if the channel is cancelled.
\begin{lstlisting}
pub fn send<T, S: Session>(x: T, s: Send<T, S>) -> S
pub fn recv<T, S: Session>(s: Recv<T, S) -> Option<(T, S)>
\end{lstlisting}
The \lstinline{close} function closes a \emph{completed} session, \ie of type \lstinline{End}. Closing is \emph{synchronous}, meaning that \lstinline{close} blocks until the dual end point is also closed. Closing a cancelled channel fails.
\begin{lstlisting}
pub fn close(s: End) -> Option<()>
\end{lstlisting}
The \lstinline{fork} function creates a new session, and spawns a new thread. The new thread runs the first argument of \lstinline{fork}---a function---which receives one endpoint, and \lstinline{fork} returns the dual endpoint. It has the following type---where \lstinline{FnOnce(S) -> Option<()>} is the type of functions from \lstinline{S} to \lstinline{Option<()>} which can be invoked only once:
\begin{lstlisting}
pub fn fork<S: Session>(p: P) -> S::Dual
where
  P: FnOnce(S) -> Option<()>,
\end{lstlisting}
The function passed to \lstinline{fork} is allowed to fail, hence the \lstinline{Option}-type. If it fails, the child thread panics silently, and all open sessions in the child thread are cancelled.

Now that we have covered the basics, let us consider an example program. The following program forks off a child process which sends a ``ping'', then waits to receive that ping and returns:
\begin{lstlisting}
let s = fork(move |s: Send<(), End>| {
  let s = send((), s);
  close(s)
});
let ((), s) = recv(s)?;
close(s)
\end{lstlisting}
The \lstinline{let x = M?; N} construct is Rust's ``monadic bind'' notation for programs which may return errors. If \lstinline{recv(s)} succeeds, the \lstinline{?}-operator unpacks the \lstinline{Option}. If \lstinline{recv(s)} fails, the \lstinline{?}-operator short-circuits, skips the rest of the statements, and returns the error.

\subsection{Linearity and cancellation}\label{sec:cancellation}
Session types often require linearity---the protocol for a negation server in~\cref{sec:basics} specifies that the server receives \emph{one} integer, and then sends \emph{one} integer back. For such a server to, \eg send back multiple integers would be, at best, confusing, and at worst, cause an error. The following two erroneous programs violate the linearity condition. The program on the left reuses the channel s, and the program on the right drops it without completing the session:

\vspace{-0.5\baselineskip}
\begin{minipage}[t]{0.5\linewidth}
\begin{lstlisting}
let s = fork(move |s: Send<(), End>| {
  let s1 = send((), s);
  close(s1)?;
  let s2 = send((), ~s~); // reuse s
  close(s2)
});
let ((), s) = recv(s)?;
close(s)
\end{lstlisting}
\end{minipage}%
\begin{minipage}[t]{0.5\linewidth}
\begin{lstlisting}
let s = fork(move |s: Send<(), End>| {
  
  // unintentionally
  // left blank.

});
let ((), s) = recv(s)?;
close(s)
\end{lstlisting}
\end{minipage}
\vspace{-0.25\baselineskip}

\noindent
Rust is an affine language, meaning that any value can be used \emph{at most once}. Hence, our first example doesn't compile: Rust complains about the second usage of \lstinline{s}, which attempts to use \lstinline{s} after transferring ownership of the value to \lstinline{send}.

The second program is trickier, as Rust's type system doesn't help us here. Arguably, however, that is for the best. The idea of \emph{linear} session types ignores the reality that sessions may be dropped at any time, for any number of reasons: a process may panic, dropping all sessions in scope, or---when communicating over the network---the connection might fail. In~\cref{sec:introduction}, we mentioned that EGV has an affine type system, and permits the explicit cancellation of sessions. Following EGV, Rusty Variation also permits the cancellation of sessions. However, instead of having cancellation be an explicit primitive, we implement it implicitly for any dropped value, using Rust's destructors\footnote{\url{https://doc.rust-lang.org/1.35.0/book/ch15-03-drop.html}}.

When an initialised value in Rust goes out of scope, Rust calls its destructor, the \lstinline{drop} method. This method can be customised by implementing the \lstinline{Drop} trait. For instance, if we use the implementation of \lstinline{Drop} shown on the left, running the code on the right would print ``End dropped!'' twice, once from the child thread, and once from the main thread.

\vspace{-0.5\baselineskip}
\begin{minipage}[t]{0.5\linewidth}
\begin{lstlisting}
impl Drop for End {
  fn drop(&mut self) {
    println!("End dropped!")
  }
}
\end{lstlisting}
\end{minipage}%
\begin{minipage}[t]{0.5\linewidth}
\begin{lstlisting}
let _s = fork(move |_s: End| {
  // intentionally left blank.
});
// intentionally left blank.
\end{lstlisting}
\end{minipage}
\vspace{-0.25\baselineskip}

\noindent
To mimic the cancellation behaviour in EGV, we count the number of references to a session, starting at \emph{two} when a session is created using \lstinline{new}. Each time a session's destructor is called, we decrement this counter. If the counter is \emph{one}, the session is marked as \emph{disconnected}. Sending on a disconnected session drops the value being sent. Receiving on or closing disconnected sessions returns an error, \ie \lstinline{None}. This ensures that no process is left waiting for a message that will never come. If the counter reaches \emph{zero}, the memory for the session is deallocated.

As it turns out, this is exactly the behaviour for the channels from \lstinline{crossbeam_channel}, which we use in our implementation, and when a session is dropped, the channel it wraps is dropped recusively. Hence, we inherit the correct behaviour of dropping sessions from \lstinline{crossbeam_channel}.

However, there is one major downside to replacing explicit cancellation with implicit cancellation: it is no longer possible to tell whether the programmer wanted to cancel a session or simply forgot to complete the session. We do two things to remedy this. First, the definitions of \lstinline{Send}, \lstinline{Recv}, and \lstinline{End} are annotated with \lstinline{#[must_use]}. This causes Rust to emit a warning whenever a session is dropped. Second, we provide an explicit \lstinline{cancel} function, though its implementation is simply to drop its argument:
\begin{lstlisting}
  pub fn cancel<T>(_x: T) -> () { /* intentionally left blank. */ }
\end{lstlisting}

The following two programs illustrate our solution. The program on the left forgets to use \lstinline{s}, and Rust emits a warning, complaining that \lstinline{s} isn't used. The program on the right passes the session to \lstinline{cancel}, explicitly marking the drop, and hence compiles without any warnings. Both programs have the same semantics, though: the program forks off a process which is \emph{expected} to send a ``ping''. Instead the forked process cancels the session---be it implicitly or explicitly. Hence, the call to \lstinline{recv} fails, the \lstinline{?}-operator short-circuits, and the whole program returns \lstinline{None}:

\vspace{-0.5\baselineskip}
\begin{minipage}[t]{0.5\linewidth}
\begin{lstlisting}
let s = fork(move |s: Send<(), End>| {
  // shouldn't we be using s?
  Some(())
});
let (z, s) = recv(s)?;
close(s)
\end{lstlisting}
\end{minipage}%
\begin{minipage}[t]{0.5\linewidth}
\begin{lstlisting}
let s = fork(move |s: Send<(), End>| {
  cancel(s);
  Some(())
});
let (z, s) = recv(s)?;
close(s)
\end{lstlisting}
\end{minipage}
\vspace{-0.25\baselineskip}

\subsection{Branching with offer! and choose!}\label{sec:choice}
The \lstinline{choose_left}, \lstinline{choose_right}, and \lstinline{offer_either} functions let processes offer or make a \emph{binary} choice. These are \emph{derived} constructs~\cite{honda1993,dardha2017}. They are implemented using only the functions in our library---\lstinline{new}, \lstinline{send}, \lstinline{recv}, \lstinline{close}, and \lstinline{cancel}---and the sum type \lstinline{Either}. They send or receive a value of the type \lstinline{Either<S1, S2>}, where \lstinline{S1} and \lstinline{S2} are the possible continuations of the session.
\begin{lstlisting}
type Choose<S1, S2> = Send<Either<S1, S2>, End>;
type Offer<S1, S2> = Recv<Either<S1, S2>, End>;

pub fn choose_left<S1: Session, S2: Session>(s: Choose<S1, S2>) -> S1
pub fn choose_right<S1: Session, S2: Session>(s: Choose<S1, S2>) -> S2
pub fn offer_either<S1: Session, S2: Session, P1, P2, R>(
       s: Offer<S1, S2>, p1: P1, p2: P2) -> Option<R>
where
  P1: FnOnce(S1) -> Option<R>,
  P2: FnOnce(S2) -> Option<R>,
\end{lstlisting}
There is a subtlety in their implementation. The \lstinline{choose_left} function creates a new session of type \lstinline{S1}. It wraps one endpoint in \lstinline{Either::Left}, sends it over the original session \lstinline{s}, and returns the other endpoint. However, this leaves it with the continuation of the original session, of type \lstinline{End}. Closing this endpoint synchronously, using \lstinline{close}, may fail. However, \lstinline{choose_left} \emph{sends}, and for consistency with \lstinline{send}, we would like it to always succeed. Therefore, we opt to \emph{cancel} the continuation instead:
\begin{lstlisting}
pub fn choose_left<S1: Session, S2: Session>(s: Choose<S1, S2>) -> S1 {
    let (here, there) = S1::new();
    let s = send(Either::Left(there), s);
    cancel(s);
    here
} 
\end{lstlisting}
Cancelling the endpoint is harmless---as long as the other processes is prepared, and uses the corresponding function \lstinline{offer_either}, this acts as an \emph{asynchronous} alternative to close. Pairing \lstinline{choose_left} with a manually implemented version of \lstinline{offer_either} which attempts to \emph{close} the continuation instead of cancelling it results in a runtime error.

The \lstinline{choose_left}, \lstinline{choose_right}, and \lstinline{offer_either} functions suffice to encode choice, they are not useful in practice---a protocol with a four-way choice would require three nested \lstinline{Offer} constructs, and several calls to \lstinline{offer_either}. Code written in this style rapidly becomes unreadable. For this reason, we also define two macros, \lstinline{choose!} and \lstinline{offer!}, which generalise binary choice using \lstinline{Either} to a choice on \emph{any} \lstinline{enum}, as long as each case wraps a session. This allows processes to offer and make choices between many \emph{labelled} branches, and allows us to define the type for a calculator server as:
\begin{lstlisting}
enum CalcOp { Sqr(SqrSrv), Neg(NegSrv) }
type CalcSrv = Recv<CalcOp, End>;
type CalcCli = CalcSrv::Dual;
\end{lstlisting}
Using the \lstinline{choose!} and \lstinline{offer!} macros, we can implement a calculator server and client. The \lstinline{main} function below forks off a calculator server, which offers the choice between squaring and negation. The client then selects squaring, sends the number four, and prints the result:

\begin{minipage}[t]{0.5\linewidth}
\begin{lstlisting}
fn server(s: CalcSrv) -> Option<()> {
  offer!(s, {
    CalcOp::Sqr(s) => {
      let (x, s) = recv(s)?;
      let s = send(x * x, s);
      close(s)
    },
    CalcOp::Neg(s) => {
      let (x, s) = recv(s)?;
      let s = send(-x, s);
      close(s)
    },
  })
}
\end{lstlisting}
\end{minipage}%
\begin{minipage}[t]{0.5\linewidth}
\begin{lstlisting}
fn client(s: CalcCli) -> Option<i32> {
  let s = choose!(CalcOp::Sqr, s);
  let s = send(4, s);
  let (z, s) = recv(s)?;
  close(s)?;
  Some(z)
}

// main() prints Some(16)
fn main() {
  let s = fork(server);
  let z = client(s);
  println!("{}", z);
}
\end{lstlisting}
\end{minipage}
\vspace{-0.25\baselineskip}

\subsection{Homogeneous Selection}\label{sec:selection}
Selection is a mechanism which allows processes to block on a list of actions, waiting until the first of them fires. Writing programs which use selection can be tricky and error prone, even using libraries specifically built to help with selection. For instance, the following is an excerpt from the documentation of \lstinline{crossbeam_channel::Select}\footnote{\url{https://docs.rs/crossbeam-channel/0.3.8/crossbeam_channel/struct.Select.html}}, a construct which implements a selection mechanism:
\begin{quotation}
  \noindent
  Once a list of operations has been built with \lstinline{Select}, there are two different ways of proceeding:
  \begin{itemize}
  \item Select an operation with \lstinline{try_select}, \lstinline{select}, or \lstinline{select_timeout}. If successful, the returned selected operation has already begun and \emph{must} be completed. If we don't complete it, a panic will occur.
  \item Wait for an operation to become ready with \lstinline{try_ready}, \lstinline{ready}, or \lstinline{ready_timeout}. If successful, we may attempt to execute the operation, but are not obliged to. In fact, it's possible for another thread to make the operation not ready just before we try executing it, so it's wise to use a retry loop. However, note that these methods might return with success spuriously, so it's a good idea to always double check if the operation is really ready.
  \end{itemize}
\end{quotation}
None of this advice is enforced by Rust---instead, this must be checked by the programmer---and bugs in the use of selection can be hard to find, especially if the approach using \lstinline{ready} is used.

RV offers an alternative, the \lstinline{select_mut} function. This function operates on a vector of \emph{receiving} endpoints, returning the result of the first receive operation which completes, and removing the corresponding session from the vector. We restrict ourselves to the receiving endpoints, as RV's \lstinline{send} operation is non-blocking, hence selecting on sending endpoints doesn't make any sense.
\begin{lstlisting}
pub fn select_mut<T, S: Session>(rs: &mut Vec<Recv<T, S>>) -> Option<(T, S)>
\end{lstlisting}

The \lstinline{select_mut} function is implemented on top of \lstinline{crossbeam_channel::Select}, in a way which respects the rules listed above. In addition, RV offers an immutable variant of \lstinline{select_mut}, named \lstinline{select}, which returns the remainder of the endpoints along the with result of the selection. Using \lstinline{select_mut}, we can write the following program, which forks off ten threads, each sending an index between 1 and 10, and then repeatedly selects on the receiving endpoints, receiving and printing an index, printing the numbers 1 to 10 in a random order:

\vspace{-0.5\baselineskip}
\begin{minipage}[t]{0.5\linewidth}
\begin{lstlisting}
let mut rs = Vec::new();
for i in 0..10 {
  let s = fork(move |s: Send<i32, End>| {
    let s = send(i, s);
    close(s)
  });
  rs.push(s);
}
\end{lstlisting}
\end{minipage}%
\begin{minipage}[t]{0.5\linewidth}
\begin{lstlisting}
loop {
  if rs.is_empty() {
    break Ok(());
  }
  let (i, r) = select_mut(&mut rs)?;
  println!("{}", i);
  close(r)
}
\end{lstlisting}
\end{minipage}
\vspace{-0.25\baselineskip}

The \lstinline{select_mut} and \lstinline{select} functions are homogeneous: they only select over vectors of endpoints with \emph{the same type}. This rules out heterogeneous uses of selection, where the operations are of different types. Selection is not a part of EGV, and hence our strong correctness guarantees do not extend to selection.

\subsection{Implementation}\label{sec:implementation}
So far we have avoided discussing the implementation whenever possible, and with good reason: RV can be implemented using \emph{any} library which offers one-shot channels and selection. Hence, the important contribution in this paper is not the implementation, but the programming interface offered by RV. 

Let us briefly discuss the implementation. We use the asynchronous channels from \lstinline{crossbeam_channel}, which at the time of writing is the de facto implementation of channels in Rust. Channels are typed, and can only transfer values of a single type. Therefore, we use one-shot channels, and encode sessions following Scalas and Yoshida, and Padovani~\cite{scalas2016,padovani2017}: Each \lstinline{send} creates a new session for the continuation, and uses the primitive \lstinline{send} operation to send the value along with one endpoint of the continuation session. If the primitive \lstinline{send} operation returns an error, we swallow it, using \lstinline{unwrap_or}. Each \lstinline{recv} receives a value together with a channel on which to continue the session. If the primitive \lstinline{recv} operation returns an error, the \lstinline{?}-operator short-circuits, and we return the error. The channels in \lstinline{crossbeam_channel} are highly optimised for one-shot usage.
\begin{lstlisting}
use crossbeam_channel::{Sender, Receiver}; // abbreviated import list

pub struct Send<T, S: Session> { channel: Sender<(T, S::Dual)> }
pub struct Recv<T, S: Session> { channel: Receiver<(T, S)> }

pub fn send<T, S: Session>(x: T, s: Send<T, S>) -> S
{
    let (here, there) = S::new();
    s.channel.send((x, there)).unwrap_or(());
    here
}

pub fn recv<T, S: Session>(s: Recv<T, S>) -> Option<(T, S)>
{
    let (v, s) = s.channel.recv()?;
    Some((v, s))
}
\end{lstlisting}
Closing is implemented using two asynchronous channels, which we use to simulate a single synchronous channel. Both channels transmit a unit value. Each close operator first sends on one of the channels (for which it has the sender) and then receives (blocking) on the other channel (for which it has the receiver). If the other endpoint is cancelled, both channels are marked as disconnected, and thus any close operation which is still blocking will raise an exception. Otherwise, the first close operation blocks until the second is executed.
\begin{lstlisting}
pub struct End { sender: Sender<()>, receiver: Receiver<()> }

pub fn close(s: End) -> Option<()> {
    s.sender.send(()).unwrap_or(());
    s.receiver.recv()?;
    Some(())
}
\end{lstlisting}
While the current implementation is efficient, it could likely be improved by replacing \lstinline{crossbeam_channel} with a minimal implementation of one-shot channels and selection.

The Rust standard library offers \lstinline{std::sync::mpsc}, an implementation of channels. The first implementation of RV used the channels from this library. However, \lstinline{std::sync::mpsc} has some fundamental issues with its design\footnote{\url{https://github.com/rust-lang/rust/pull/42397\#issuecomment-315867774}}, and seems to be on its way out. In 2018, the selection mechanism, \lstinline{mpsc_select!}, was deprecated\footnote{\url{https://github.com/rust-lang/rust/issues/27800}}, and there are plans to phase out \lstinline{std::sync::mpsc} altogether~\cite{glavina2019}.

\subsection{What If I Want to Be Evil?}\label{sec:limitations}
There are three ways in which you can make code written using Rusty Variation lock in safe Rust. The first is by writing an infinite loop, which we discussed earlier. 

The biggest issue lies in the fact that Rust's semantics do not guarantee that the destructors are actually run. This does not mean that Rust can \emph{arbitrarily} decide not to run destructors, but that it is possible to write programs which leak memory. The most obvious example of this is \lstinline{std::mem::forget}\footnote{\url{https://doc.rust-lang.org/1.35.0/std/mem/fn.forget.html}}:
\begin{quotation}
  \noindent
  Takes ownership and ``forgets'' about the value \emph{without running its destructor}.  

  \noindent
  Any resources the value manages, such as heap memory or a file handle, will linger forever in an unreachable state. However, it does not guarantee that pointers to this memory will remain valid.
\end{quotation}
Explicitly calling \lstinline{std::mem::forget} on session endpoints is unlikely to happen by accident, hence we don't consider this a huge shortcoming. However, it is also possible to leak memory by creating reference cycles\footnote{\url{https://doc.rust-lang.org/1.35.0/book/ch15-06-reference-cycles.html}}---in fact, \lstinline{std::mem::forget} was initially marked as \lstinline{unsafe}, until it was discovered that it could be implemented in safe Rust creating reference cycles. This means that care needs to be taken when combining sessions and cyclic data structures, \eg storing session endpoints in doubly linked lists.

The second problem arises from the availability of \lstinline{new}, which creates two dual session endpoints. Code written using only \lstinline{fork} is guaranteed to be deadlock free~\cite{lindley2015}, but using \lstinline{new} and \lstinline{std::thread::spawn}, it becomes easy to write code which deadlocks. However, it is impossible to export a trait without exporting all its members. Arguably, hiding \lstinline{new} is also undesirable. Using only \lstinline{fork}, it is impossible to create any cyclic communication structures, including those which are well behaved. Therefore, a programmer may occasionally \emph{need} access to \lstinline{new}.

The following two programs illustrate the two ways of constructing deadlocking programs discussed above. The program on the left uses \lstinline{std::mem::forget} to drop the session endpoint without running its destructors. The program on the right creates two pairs of channels to construct a deadlock:
\vspace{-0.5\baselineskip}
\begin{minipage}[t]{0.5\linewidth}
\begin{lstlisting}
// Deadlock using std::mem::forget
let s = fork(move |s: Send<(), End>| {
  std::mem::forget(s);
  Some(())
});
let ((), s) = recv(s)?;
close(s)
\end{lstlisting}
\end{minipage}%
\begin{minipage}[t]{0.5\linewidth}
\begin{lstlisting}
// Deadlock using new and spawn
let (s1, r1) = Send<Void, End>::new();
let (s2, r2) = Send<Void, End>::new();
std::thread::spawn(move || {
  let (v, r1) = recv(r1)?;
  close(r1)?;
  let s2 = send(v, s2);
  close(s2)
});
let (v, r2) = recv(r2)?;
close(r2)?;
let s1 = send(v, s1);
close(s1)
\end{lstlisting}
\end{minipage}
\vspace{-0.25\baselineskip}

\section{Related Work}\label{sec:related-work}
Laumann \etal~\cite{jespersen2015} implement session types in Rust, based on the session-typed functional language LAST by Gay and Vasconcelos~\cite{gay2009}. Their implementation guarantees session fidelity, albeit with some caveats. LAST is a linear language, meaning that sessions can neither be duplicated nor dropped. As we discussed in~\cref{sec:cancellation}, Rust is an affine language, meaning that values---and, particularly relevant here, sessions---can be dropped. The implementation of LAST in Rust, therefore, guarantees session fidelity under the assumption that channels are \emph{never dropped}.

Laumann \etal~\cite{jespersen2015} remark that they ``cannot prevent a channel value from being dropped prematurely''. They offer several solutions, but remark that each of these solutions are either insufficient, premature, or both. They note that the proposal to add linear types to Rust is currently postponed\footnote{\url{https://github.com/rust-lang/rfcs/pull/776}}, and that while \lstinline{#[must_use]} can be made to issue a warning if a value is dropped, it is ``easy to accidentally and purposely bypass''. Finally, they propose a compiler extension, developed in collaboration with Manish Goregaokar, \lstinline{humpty_dumpty}\footnote{\url{https://github.com/Manishearth/humpty_dumpty}}, ``which provides a lint that ensures that types annotated  with \lstinline{#[drop_protect]} are used linearly''. However, they note that \lstinline{humpty_dumpty} ``is still a work in progress, which by no means guarantees linearity in all cases''. No further work has been done on \lstinline{humpty_dumpty} since 2015. The current version of the library enforces linearity \emph{dynamically}, using ``destructor bombs'', implemented by Manish Goregaokar\footnote{\url{https://github.com/Munksgaard/session-types/commit/0f25ccb7c3bc9f65fa8eaf538233e8fe344a189a}}. This technique hides a \lstinline{panic!} in channel destructors, which crashes the program if the destructor ever runs. Each invocation of \lstinline{close} then calls \lstinline{std::mem::forget}, bypassing the destructor bomb.

We argue that even if Rust were able to enforce linearity statically, the \emph{idea} of linear session types is flawed (see~\cref{sec:cancellation}), as sessions may be dropped at any time, for any number of reasons, \eg a process panicking, dropping all sessions in scope.

The following two programs illustrate the difference in handling failure between Rusty Variation and the work by Laumann \etal~\cite{jespersen2015}. In both programs, the programmer has forgotten to complete the server, accidentally dropping the session endpoint \lstinline{_s} or channel \lstinline{_c}, respectively. The program on the left, written in RV, returns \lstinline{None}. The program on the right, written using the library described by Laumann \etal\cite{jespersen2015}, segfaults. We have since reported this segfault to Laumann \etal, and they have fixed it in the most recent version of the library. The program on the right now panics:

\vspace{-0.5\baselineskip}
\begin{minipage}[t]{0.5\linewidth}
\begin{lstlisting}
// Rusty Variation
// main() returns None
fn server(_s: Recv<(), End>) {
  // unintentionally left blank.
}
fn client(s: Send<(), End>) {
  close(send((), s))
}
fn main() -> Option<()> {
  client(fork(server))
}
\end{lstlisting}
\end{minipage}%
\begin{minipage}[t]{0.5\linewidth}
\begin{lstlisting}
// Laumann et al.
// main() causes segfault
fn server(_c: Chan<(), Recv<(), Eps>>) {
  // unintentionally left blank.
}
fn client(c: Chan<(), Send<(), Eps>>) {
  c.send(()).close();
}
fn main() -> () {
  connect(server, client);
}
\end{lstlisting}
\end{minipage}
\vspace{-0.25\baselineskip}

While we haven't discussed recursive session types in this paper, Laumann \etal~\cite{jespersen2015} use a rather cumbersome encoding for recursive session types, using type-level de Bruijn indices. Below we show two definitions of the session type for a stream of integers, one in Rusty Variation, and one using the library by Laumann \etal~\cite{jespersen2015}:

\begin{minipage}{0.5\linewidth}
\begin{lstlisting}
// Rusty Variation
type IntStream = Send<i32, IntStream>;
\end{lstlisting}  
\end{minipage}%
\begin{minipage}{0.5\linewidth}
\begin{lstlisting}
// Laumann et al.
type IntStream = Rec<Send<i32, Var<Z>>>;
\end{lstlisting}  
\end{minipage}
\vspace{-0.25\baselineskip}

\noindent
The usage of de Bruijn indices to encode recursive session types is an artefact of the fact that, when Laumann \etal~\cite{jespersen2015} implemented their library, Rust didn't support recursive types.

A last feature which distinguishes Rusty Variation from the work by Laumann \etal~\cite{jespersen2015} is in how both libraries generalise binary choice to labelled choice. Rusty Variation uses enumerations to provide labelled choice, as described in~\cref{sec:choice}. Laumann \etal~\cite{jespersen2015} provide an \lstinline{offer!} macro, which expands a labelled offer to a series of binary offers, \eg offering a ternary choice expands to two nested invocations of \lstinline{offer_either}.

The \lstinline{offer!} macro allows---even requires---the programmer to label their choices, but during macro expansion these labels are simply discarded \emph{without being checked}. This allows us to write the program below, in which an inconsistency between types and documentation on the one hand, and the labelled implementation on the other hand, results in a bug. The example program implements a calculator server, which offers the choice between squaring and negation, in that order.

In the \lstinline{server} function, the order in which the \lstinline{NEG}-case and the \lstinline{SQR}-case are handled doesn't match the order in which they are declared in the types. However, the code compiles, not because the cases are associated correctly through the use of labels, but because the labels aren't checked, and squaring and negation happen to have the same type signature.

When the client selects the first option, they wrongly assume it represents squaring, based on the types. Unfortunately, it represents negation. The \lstinline{main} program, when run, therefore execute negation, and erroneously returns \lstinline{-42}:

\vspace{-0.5\baselineskip}
\begin{minipage}[t]{0.5\linewidth}
\begin{lstlisting}
// Laumann et al.
type SqrSrv  = Recv<i32, Send<i32, Eps>>;
type NegSrv  = Recv<i32, Send<i32, Eps>>;
type CalcSrv = Offer<SqrSrv, NegSrv>;
\end{lstlisting}
\end{minipage}%
\begin{minipage}[t]{0.5\linewidth}
\begin{lstlisting}

type SqrCli  = Send<i32, Recv<i32, Eps>>;
type NegCli  = Send<i32, Recv<i32, Eps>>;
type CalcCli = Choose<SqrCli, NegCli>;
\end{lstlisting}
\end{minipage}
\vspace{-0.25\baselineskip}

\vspace{-0.5\baselineskip}
\begin{minipage}[t]{0.5\linewidth}
\begin{lstlisting}
fn server(c: Chan<(), CalcSrv>) {
  offer!{c,
    NEG => {
      let (c, n) = c.recv();
      let c = c.send(-n);
      c.close();
    },
    SQR => {
      let (c, n) = c.recv();
      let c = c.send(n * n);
      c.close();
    }
  }
}
\end{lstlisting}
\end{minipage}%
\begin{minipage}[t]{0.5\linewidth}
\begin{lstlisting}
fn client(c: Chan<(), CalcCli>) {
  let c = c.sel1(); // select SqrCli
  let c = c.send(42);
  let (c, n) = c.recv();
  println!("{}", n);
  c.close();
}



// main() prints -42
fn main() {
  connect(server, client);
}
\end{lstlisting}
\end{minipage}
\vspace{-0.25\baselineskip}

\section{Future Work}\label{sec:future-work}

\subsection{Heterogeneous Selection}
In~\cref{sec:selection}, we mention that the \lstinline{select_mut} and \lstinline{select} functions are homogeneous: they only select over vectors of endpoints with \emph{the same type}. Heterogeneous selection is an important operation for channel-based communication. \emph{Typing} heterogeneous selection, and by extension, heterogeneous vectors is notoriously cumbersome without dependent types. There are several avenues for future research:

The \lstinline{frunk} crate\footnote{\url{https://beachape.com/frunk/frunk/index.html}} provides an encoding of heterogeneous lists in Rust. We could use this encoding to define a heterogeneous variant of selection.

Rust has limited support for existential types\footnote{\url{https://github.com/rust-lang/rfcs/blob/master/text/2071-impl-trait-existential-types.md}}, which could be used to quantify over the types of the values received and the session continuations, and select on a homogeneous vector of the (purely fictional) type \lstinline[mathescape=true]{Vec<$\exists$T$.\exists$S$.$Recv<T,S>>}.

\subsection{Testing \& Verification}\label{sec:verification}
In~\cref{sec:contribution}, we claim that RV preserves the metatheory of EGV, and while a full discussion of the testing and formal verification efforts we have undertaken to ensure the correctness of RV is out of the scope of this paper, it would be remiss of us not substantiate this claim.

Firstly, RV is unit tested. We test various use-cases from EGV, from basic operations and their interactions with cancellation, to delegation and recursive session types. This test suite has caught several errors in the implementation. However, unit tests are error prone, and it is hard (if not impossible) to obtain a set of tests which cover all use-cases of a library. Unfortunately, tools such as RustBelt~\cite{jung2017} and Rust Distilled~\cite{weiss2018} are not yet mature enough to allow us to \emph{prove} the correctness of RV.

In the absence of a mature formal semantics for Rust, we are designing a formal language, RV (or ``formal RV''), which matches, as close as possible, the semantics of our Rust library. It is a concurrent lambda calculus with heap-based shared-memory semantics. We plan to prove that formal RV guarantees preservation, progress, termination, confluence and the diamond property, and deadlock and livelock freedom. Furthermore, we plan to show that there exists a translation from EGV to formal RV, based on the monadic reflection of exceptions into the \lstinline{Option} monad~\cite{filinski1994} and the translation of channel-based communication to shared memory, which is in operational correspondence with EGV. This would show that formal RV is a faithful implementation of EGV.

Lastly, we plan to ensure the correspondence between formal RV and the implementation of RV, using property-based testing. We plan to generate a series of random formal RV terms, using Neat~\cite{claessen2015}, and check if their behaviour, when run as Rust programs, corresponds to the behaviour of the term in formal RV.

\subsection{Extending Exceptional GV}
Several features of Rusty Variation are not covered by the formal language EGV. In~\cref{sec:selection}, we mention that selection isn't a part of EGV. In~\cref{sec:related-work}, we mention that recursion isn't part of EGV. If we wish to have any hope of formally verifying RV, along the lines described in~\cref{sec:verification}, we would need to extend EGV to cover these features.

\bibliographystyle{eptcs}
\bibliography{main}

\end{document}